\documentclass{aa}

\usepackage{graphics}

\begin{document}

\thesaurus{06(02.01.1; 
08.14.1; 08.16.6)}

\title{ Approximate expressions for polar gap electric field\\ of pulsars}

\author{J. Dyks \and B. Rudak}
\institute{Nicolaus Copernicus Astronomical Center, Rabia\'nska 8, 
87-100 Toru\'n, Poland (jinx@ncac.torun.pl; bronek@camk.edu.pl)}

\date{Received / accepted }
\titlerunning{Polar gap electric field}
\authorrunning{Dyks \& Rudak}
\maketitle

\begin{abstract}

We derive easy-to-handle approximations for the polar gap electric field
due to inertial frame dragging 
as derived by Harding \& Muslimov (\cite{hm}).
A formula valid for polar gap height comparable to the polar cap radius
is presented. 


\end{abstract}

\section{Introduction}

Accurate models for $E_\parallel$ --- the electric field component 
parallel to the local magnetic field $\vec B$ 
above polar caps of rotating magnetized neutron stars ---
are of great theoretical interest 
in the context of magnetospheric activity of pulsars.
In the framework of space charge limited flow model
(originally introduced by Arons \& Scharlemann (\cite{as})), 
Harding \& Muslimov (\cite{hm}) (hereafter HM98)
derived $E_\parallel$ including the general relativistic 
frame dragging effect, worked out by Muslimov \& Tsygan
(\cite{mt}) (hereafter MT92).
HM98 considered the case with  $E_\parallel$ screened at both a lower and 
an upper boundary of acceleration region
i.e.~at the star surface and at a pair formation front.
Since the full expression for $E_\parallel$ is too cumbersome for 
practical applications, HM98 offered simple analytic expressions
valid in various limiting cases. 
In this paper we revise and extend their results.
In particular, we show that
in the most frequently considered case, i.e. when the gap accelerator length
is comparable to the size of the polar cap radius, the approximation derived by HM98
overestimates $E_\parallel$ by a ratio of the neutron star radius to the polar cap radius.

\section{Formulae for $E_\parallel$}

In the following, $h$ denotes the altitude above the neutron star surface
and $h_c$ is the gap height above which a pair formation front
screens $E_\parallel$ (i.e.~$E_\parallel=0$ for $h\ge h_c$). 
Altitudes $z\equiv h/R_{\rm ns}$, $z_c\equiv h_c/R_{\rm ns}$ 
and radial distances
$\eta\equiv 1 + z$, $\eta_c\equiv 1 + z_c$ 
scaled with the star radius $R_{\rm ns}$
will also be used. The magnetic colatitude $\xi\equiv \theta/\theta(\eta)$ 
is scaled with the
half-opening angle of the polar magnetic flux tube $\theta(\eta)$. 


A solution of Poisson's equation for the polar gap with $h_c \ll R_{\rm ns}$
as derived by HM98 reads 

\begin{eqnarray}
\lefteqn{E_\parallel^{(1)} \simeq -E_0\ \theta_0^3\ (1 - \epsilon)^{1/2} \left\{
\frac{3}{2}\ \kappa\ S_1 \cos{\chi} \ + \right.}\nonumber\\
&&\hbox{\hskip25mm}\left.+\ \frac{3}{8}\ \theta_0 H(1)\delta(1) S_2 \sin{\chi}\cos{\phi} \right\}
\label{elf1}
\end{eqnarray} 
where,

\begin{equation}
\begin{array}{l}
S_1 = \sum\limits_{i = 1}^\infty \frac{8 J_0(k_i \xi)}{k_i^4
J_1(k_i)}
{\cal F}(z, z_c, \gamma_i),\phantom{\frac{\frac{1}{2}}{\frac{1}{4}}}\\
\\
S_2 =
\sum\limits_{i = 1}^\infty \frac{16 J_1(\tilde k_i \xi)}
{\tilde k_i^4 J_2(\tilde k_i)} {\cal F}(z, z_c, \tilde \gamma_i),   
\end{array}
\end{equation}

\begin{eqnarray}
\lefteqn{ {\cal F}(z, z_c, \gamma) = 
-[a_1(\gamma\eta - 1)e^{\gamma z} + 
a_2(\gamma\eta + 1)e^{-\gamma z}\ + }\nonumber\\
&&\hbox{\hskip20mm}+\ a_1(1 - \gamma) - a_2(1 + \gamma)]/(a_1 + a_2),
\label{calf}
\end{eqnarray}

\begin{equation}
a_1 = (\gamma\eta_c + 1)e^{-\gamma z_c} - \gamma - 1,\ \ \ 
a_2 = \gamma - 1 - (\gamma\eta_c -1 )e^{\gamma z_c}
\end{equation}
and

\begin{equation}
\gamma_i \approx \frac{k_i}{\theta_0(1 - \epsilon)^{1/2}},\ \ \ \ {\rm and}\ \ \ \ 
\tilde \gamma_i \approx \frac{\tilde k_i}{\theta_0(1 - \epsilon)^{1/2}},
\label{gi}
\end{equation}
where $k_i$ and $\tilde k_i$ are the positive roots of the Bessel functions
$J_0$ and $J_1$, respectively, $H(1)\delta(1)\approx 1$, and $\chi$ is a
tilt angle between the rotation and the magnetic dipole axes.
Values of $E_0$, $\theta_0$, $\epsilon$, and
$\kappa$ are given by:


\begin{equation}
\begin{array}{l}
E_0 \equiv B_{\rm pc}\frac{\Omega R_{\rm ns}}{c},\hbox{\hskip10mm}
\theta_0=\left( \frac{\Omega R_{\rm ns}}{c f(1)} \right)^{1/2},\\
\\
\epsilon \equiv \frac{2 G M}{R_{\rm ns} c^2},\hbox{\hskip20mm}
\kappa \equiv \frac{\epsilon I}{M R_{\rm ns}^2},
\end{array}
\end{equation}
where $B_{\rm pc}$ is the magnetic field strength at the magnetic pole, 
$\Omega$ is the pulsar rotation frequency, and
$I$, $M$ are the moment of inertia, and mass of the
neutron star, respectively.
The function $f(\eta)$  reads

\begin{equation}
f(\eta) = -3 x^3 \left[\ln{(1 - x^{-1})} + (1 + (2x)^{-1})/x\right], 
\end{equation}
with $x=\eta/\epsilon$.

The electric field $E_\parallel^{(1)}$ as a function of height $h$ is drawn 
in Fig.~1 
for several values of gap height $h_c$. It can be seen that $E_\parallel$
saturates in a twofold way:
First, for a fixed $h_c$ it assumes a constant value at $h$ exceeding the
polar cap radius $r_{\rm pc}$
(after a linear increase with $h$).
Second, for a fixed $h\ll h_c$ it initially increases linearly with $h_c$
to become constant for $h_c\mathrel{\hbox{\rlap{\hbox{\lower4pt\hbox{$\sim$}}}\hbox{$>$}}} r_{\rm pc}$.
As noted by Harding \& Muslimov this behaviour can be reproduced 
with simple approximations of
Eq.~(\ref{elf1}) for different regimes of validity.

\begin{figure}
\resizebox{\hsize}{!}{\includegraphics{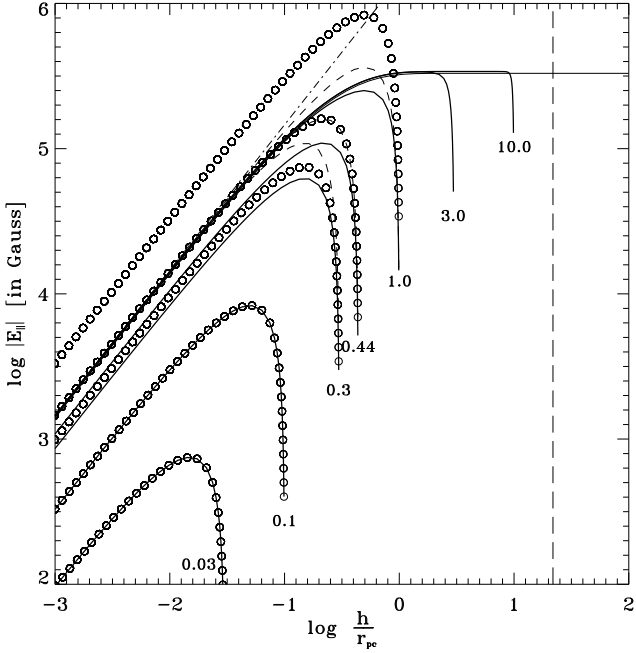}}
\caption{The electric field $E_\parallel$ in the limit $h_c \ll R_{\rm ns}$ 
as a function of height $h$ in units of $r_{\rm pc}$ 
for several values of gap height 
$h_c = 0.03$, $0.1$, $0.3$, $0.44$, $1.0$, $3.0$, and $10.0\ r_{\rm pc}$. 
Thick solid lines present $E_\parallel^{(1)}$ of Eq.~(\ref{elf1}).
Open circles mark 
approximation $E_\parallel^{(2)}$ (Eq.~\ref{app2} for 
the cases $0.03$, $0.1$, $0.3$, $0.44$,
and $1.0\ r_{\rm pc}$). The dot-dashed line is for $E_\parallel^{(4)}$
(Eq.~\ref{app4}) and dashed lines are for $E_\parallel^{(6)}$
(Eq.~\ref{app6}, the cases $h_c/r_{\rm pc}=0.3$, $0.44$, and $1.0$).
The thin solid line indicates the case with no upper
gap boundary ($E_\parallel^{(3)}$ of Eq.~\ref{app3}). 
The vertical dashed line marks $h=R_{\rm ns}$.
For $h_c=0.44\ r_{\rm pc}$, approximations $E_\parallel^{(2)}$ 
and $E_\parallel^{(6)}$ 
coincide.
We assumed
$B_{\rm pc} = 10^{12}$ G, $P= 0.1$ s, $\chi=0.2$ rad,
$\xi=0.7$, $R_{\rm ns} = 10^6$ cm, $M=1.4 M_\odot$.
}
\label{fig1}
\end{figure}

Since $k_i\ge 2.4$ it follows form Eq.~(\ref{gi}) that $\gamma_i z$ becomes
smaller than 1 for $h \la r_{\rm pc}/3$. Thus, in the limit $h \ll r_{\rm pc}/3$
and $h_c \ll r_{\rm pc}/3$
(or $\gamma_i z \ll 1$ and $\gamma_i z_c \ll 1$) the function ${\cal F}(z, z_c, \gamma_i)$
may be approximated with 

\begin{equation}
{\cal F}(z, z_c, \gamma_i) \simeq \frac{1}{2} \gamma_i^3 \left(1 -
\frac{z}{z_c}\right) z z_c
\end{equation}
which leads to 

\begin{eqnarray}
\lefteqn{E_\parallel^{(2)} \simeq -3 \frac{\Omega R_{\rm ns}}{c} \frac{B_{\rm pc}}{1 - \epsilon}
  \left(1 - \frac{z}{z_c}\right) z z_c \left[\kappa \cos{\chi}\ +
\phantom{\frac{1}{2}} \right.}\nonumber\\
&&\hbox{\hskip28mm}\left. +\ \frac{1}{2} \theta_0 \xi H(1)\delta(1)\sin{\chi}\cos{\phi}\right],
  \label{app2}
\end{eqnarray}
(HM98), where the relations

\begin{equation}
\sum\limits_{i = 1}^\infty \frac{2 J_0(k_i \xi)}
{k_i J_1(k_i)} = 1,\ \ \ {\rm and}\ \ \ 
\sum\limits_{i = 1}^\infty \frac{2 J_1(\tilde k_i \xi)}
{\tilde k_i J_2(\tilde k_i)} = \xi
\end{equation}
have been used (see eg.~MT92). As can be seen in Fig~1, in practice,
Eq.~(\ref{app2}) (open circles) reproduces Eq.~(\ref{elf1}) (thick solid line) 
for $h_c \la r_{\rm pc}/3$ (and $h \le h_c$).

Approximate behaviour of $E_\parallel$ 
for $h_c \sim r_{\rm pc}$ can be determined by considering the opposite limiting
case for the accelerator height: $h_c \gg r_{\rm pc}/3$. Assuming
$\gamma_i z_c \gg 1$, $h < (h_c - r_{\rm pc})$, (and $h_c \ll R_{\rm ns}$) 
in Eq.~(\ref{elf1}) one obtains

\begin{equation}
{\cal F}(z, z_c, \gamma_i) \simeq \gamma_i \left( 1 - e^{-\gamma_i z}\right),
\label{ef2}
\end{equation}
which results in the same formula for $E_\parallel$ as the one derived 
by MT92 for 
the case with no
upper gap boundary (with the condition $E_\parallel$ = 0
fulfilled only at the star surface):

\begin{eqnarray}
\lefteqn{E_\parallel^{(3)} \simeq 
-3 E_0 \theta_0^2 \left\{ \kappa \cos{\chi} \sum\limits_{i=1}^{\infty}
\frac{4 J_0(k_i \xi)}
{k_i^3 J_1(k_i)} \left[1 - e^{-\gamma_i z}\right] + \right.} \nonumber \\
&&\hbox{\hskip0mm}\left.+\ \theta_0 H(1)\delta(1)\sin{\chi}\cos{\phi}
\sum\limits_{i=1}^{\infty} \frac{2 J_1(\tilde k_i \xi)}
{\tilde k_i^3 J_2(\tilde k_i)}
\left[1 - e^{-\tilde \gamma_i z}\right]
\right\}
\label{app3}
\end{eqnarray}
Nevertheless, for $r_{\rm pc} \ll h_c \ll R_{\rm ns}$,
Eq.~(\ref{app3}) 
reproduces $E_\parallel^{(1)}$ (Eq.~\ref{elf1}) almost over the entire
acceleration length 
(see the thin solid line in Fig.~1)
except its very final part, where $h \mathrel{\hbox{\rlap{\hbox{\lower4pt\hbox{$\sim$}}}\hbox{$>$}}} (h_c - r_{\rm pc})$. 

Taylor expansion of (\ref{ef2}) reveals the linear increase of
$E_\parallel$ with $h$ for $h \la 0.1r_{\rm pc}$ and the saturation above $h \simeq r_{\rm pc}$:

%
\begin{eqnarray}
\lefteqn{E_\parallel^{(4)} \simeq -3 \frac{\Omega R_{\rm ns}}{c} \frac{B_{\rm pc}}{(1 - \epsilon)^{1/2}}\ \theta_0
z \left[ \kappa f_1(\xi) \cos{\chi}\ +\phantom{\frac{1}{4}}\right.}\nonumber\\
&&\hbox{\hskip25mm}\left.+\ \frac{1}{4} \theta_0 f_2(\xi) H(1)\delta(1)\sin{\chi}\cos{\phi}\right],
\label{app4}
\end{eqnarray}
for $h \la 0.1r_{\rm pc}$ and $0.5 r_{\rm pc} \la h_c \ll R_{\rm ns}$, and

\begin{eqnarray}
\lefteqn{E_\parallel^{(5)} \simeq -\frac{3}{2} \frac{\Omega R_{\rm ns}}{c} B_{\rm pc} \theta_0^2 
(1 - \xi^2)
\left[ \kappa \cos{\chi}\ +\phantom{\frac{1}{4}}\right.}\nonumber\\
&&\hbox{\hskip25mm}\left.+\ \frac{1}{4} \theta_0 \xi H(1)\delta(1)\sin{\chi}\cos{\phi}\right],
\label{app5}
\end{eqnarray}
for $r_{\rm pc} \la h \la (h_c - r_{\rm pc})$ and $2r_{\rm pc} \la h_c \ll R_{\rm ns}$. 
The magnetic colatitude profiles

\begin{equation}
f_1(\xi) = \sum\limits_{i = 1}^\infty \frac{4 J_0(k_i \xi)}
{k_i^2 J_1(k_i)},\ \ \ {\rm and}\ \ \ f_2(\xi) = \sum\limits_{i = 1}^\infty
\frac{8 J_1(\tilde k_i \xi)}
{\tilde k_i^2 J_2(\tilde k_i)} 
\label{profiles}
\end{equation}
are shown in Fig.~2 and the relations

\begin{equation}
\sum\limits_{i = 1}^\infty \frac{8 J_0(k_i \xi)}
{k_i^3 J_1(k_i)} = 1 - \xi^2,\ \ \ 
\sum\limits_{i = 1}^\infty \frac{16 J_1(\tilde k_i \xi)}
{\tilde k_i^3 J_2(\tilde k_i)} = \xi(1 - \xi^2)
\end{equation}
have been used in derivation of (\ref{app5}) (eg.~MT92).
Approximation (\ref{app4}) is shown in Fig.~1 as
a dot-dashed line.

For $h_c \ll R_{\rm ns}$ the symmetry

\begin{equation}
E_\parallel^{(1)}(h) \simeq E_\parallel^{(1)}(h_c - h)
\label{symm}
\end{equation}
may easily be proven to hold. Therefore,
to extend the validity of Eq.~(\ref{app4}) to $h\le h_c$ it is useful to
construct the approximation

\begin{equation}
E_\parallel^{(6)} = E_\parallel^{(4)} \cdot \left(1 - \frac{z}{z_c}\right)
\label{app6}
\end{equation}
which works reasonably well for $0.5r_{\rm pc} \la h_c \la r_{\rm pc}$ (dashed lines in Fig.~1).

\begin{figure}
\resizebox{\hsize}{!}{\includegraphics{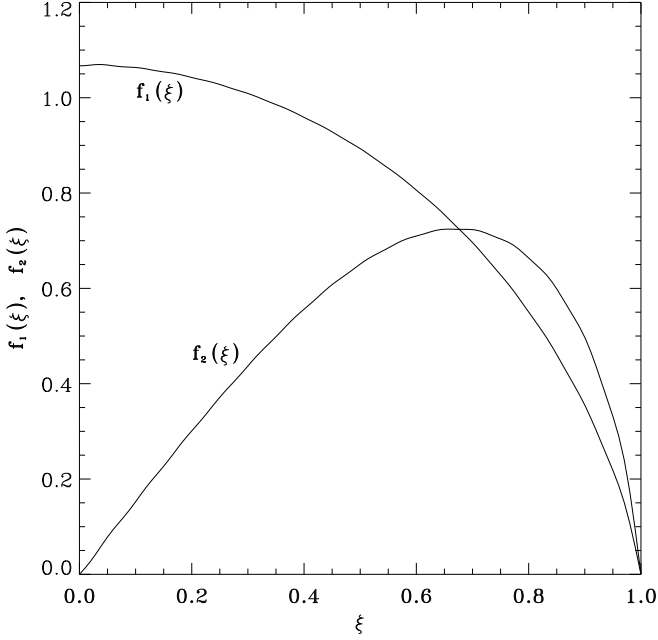}}
\caption{Magnetic colatitude 
functions $f_1(\xi)$ and $f_2(\xi)$ (see Eq.~(\ref{profiles}) for
definition).
}
\label{fig2}
\end{figure}

The expression derived by HM98 for the case $h_c \sim r_{\rm pc}$
clearly overestimates $E_\parallel$ by a factor 
$\theta_0^{-1} \approx (r_{\rm pc}/R_{\rm ns})^{-1}$
(cf. eq.~(A3) in HM98, also eq.~(1) in Harding \& Muslimov (\cite{hma}), 
or a formula used by Zhang, Harding \& Muslimov
(\cite{zhm})).
Moreover, unlike in eq.~(A3) of HM98, $E_\parallel^{(4)}$ depends
on the magnetic colatitude $\xi$ via $f_1(\xi)$. 
Our approximations are correct because
they reproduce the ``exact" formula for $E_\parallel^{(1)}$ (Eq.~\ref{elf1}).
Moreover, $E_\parallel^{(4)}$ of Eq.~(\ref{app4}) 
is identical to the formula of MT92
for the case with no screening at $h_c$
(taken in the limit $\gamma_i z \ll 1$).
This identity should hold
since for $r_{\rm pc} < h_c \ll R_{\rm ns}$ both gap boundaries influence $E_\parallel$ only within
corresponding adjacent regions of height $\sim r_{\rm pc}$.

Since Eq.~(\ref{symm}) holds for 
$h_c \ll R_{\rm ns}$,
$E_\parallel$ in the upper part of accelerator ($r_{\rm pc} < h \le h_c$) may be approximated with

\begin{equation}
E_\parallel^{(7)}(h) \simeq E_\parallel^{(3)}(h_c - h) 
\simeq \left\{
\begin{array}{l}
E_\parallel^{(4)}(h_c - h),\\
\hbox{\hskip0mm} {\rm for}\ (h_c - r_{\rm pc}) \la h \le h_c\\
\\
E_\parallel^{(5)},\\
\hbox{\hskip0mm}{\rm for}\ r_{\rm pc} < h \la (h_c - r_{\rm pc}).
\end{array}
\right.
\label{app7}
\end{equation}

In the limit where $h_c \gg r_{\rm pc}$ and $h \gg r_{\rm pc}$,
a solution of Poisson's equation 
for the elongated polar gap reads

\begin{eqnarray}
\lefteqn{E_\parallel^{(8)} \simeq -E_0 \theta_0^2\left\{
\frac{3}{2}\frac{\kappa}{\eta^4} \cos{\chi} \left[
\vphantom{\left(\frac{\eta_c}{\eta}\right)^4
   \sum\limits_{i=1}^{\infty}\frac{8 J_0(k_i \xi)}
   {k_i^3 J_1(k_i)}}
(1 - \xi^2) \ + \right.\right.}\nonumber\\
&&\left. -\ \left(\frac{\eta_c}{\eta}\right)^4
  \sum\limits_{i=1}^{\infty}\frac{8 J_0(k_i \xi)}
  {k_i^3 J_1(k_i)} e^{-\gamma_i(\eta_c)(\eta_c -
  \eta)}\right]\ +\nonumber\\
\label{elf2}
&&+\ \frac{3}{8}\ g(\eta) \sin{\chi}\cos{\phi} \left[
    \vphantom{
      \frac{\eta_c}{\eta}
      \frac{g(\eta_c)}{g(\eta)}
      \sum\limits_{i = 1}^\infty \frac{16 J_1(\tilde k_i \xi)} 
      {\tilde k_i^3 J_2(\tilde k_i)}
             }
      \xi(1 - \xi^2) \ +\right.\nonumber\\
&&\left.\left.-\ \frac{\eta_c}{\eta}
\frac{g(\eta_c)}{g(\eta)}\sum\limits_{i = 1}^\infty \frac{16 J_1(\tilde k_i
\xi)} {\tilde k_i^3 J_2(\tilde k_i)} 
e^{-\tilde \gamma_i(\eta_c)(\eta_c - \eta)}\right]\right\}
\end{eqnarray}
(MT92), where now 


\begin{equation}
\begin{array}{l}
\gamma_i \approx \frac{k_i}{\theta(\eta_c)\eta_c(1 - \epsilon/\eta_c)^{1/2}},\\
\\
\tilde \gamma_i \approx \frac{\tilde k_i}{\theta(\eta_c)\eta_c(1 - \epsilon/\eta_c)^{1/2}},
\end{array}
\end{equation}
$g(\eta) = \theta(\eta)\delta(\eta)H(\eta)$, $\theta(\eta) = \theta_0 (\eta
f(1)/f(\eta))^{1/2}$, and the functions $\delta$, and $H$ are defined 
in MT92 or HM98.
Since $\gamma_i\gg 1$, 
the two summation terms which reflect the screening effect of 
the pair formation front at $h_c$ contribute to $E_\parallel$ only at $h$ very
close to $h_c$ (see Fig.~3) so that $\gamma_i(\eta_c)(\eta_c - \eta) \sim 1$.
Thus, for $h_c \gg r_{\rm pc}$  and $r_{\rm pc} \ll h \le h_c$ one may simply   
use a formula derived in MT92 for the case with no screening at $h_c$:

\begin{eqnarray}
\lefteqn{E_\parallel^{(9)} \simeq -E_0 \theta_0^2\left\{
\frac{3}{2}\frac{\kappa}{\eta^4}
(1 - \xi^2) \cos{\chi}\ +\right.}\nonumber\\
&&\hbox{\hskip28mm}\left.+\ \frac{3}{8} g(\eta)\xi(1 - \xi^2) 
\sin{\chi}\cos{\phi}\right\}.
\label{app9}
\end{eqnarray}

\begin{figure}
\resizebox{\hsize}{!}{\includegraphics{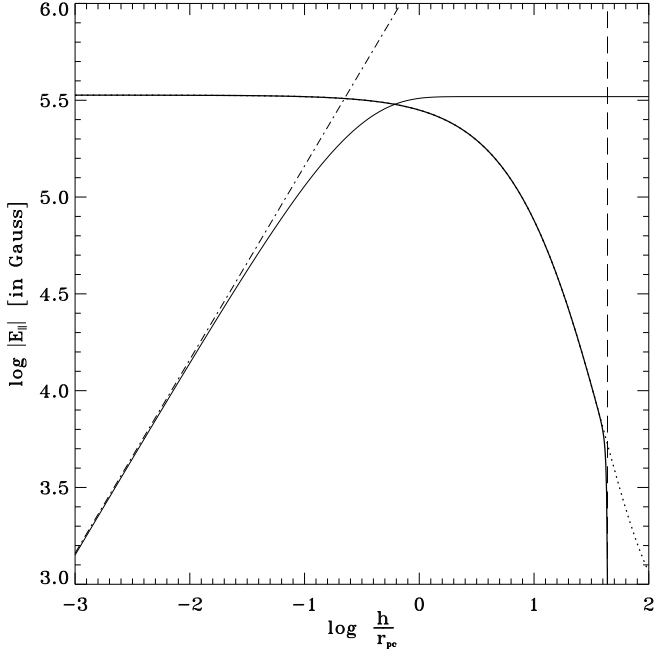}}
\caption{The run of $E_\parallel$ 
as a function of $h/r_{\rm pc}$
in the limit $h_c \gg r_{\rm pc}$.
We choose $h_c=2R_{\rm ns}$ (marked with dashed vertical line); the other parameters 
are assumed as in Fig.~1. 
Thick solid line presents $E_\parallel^{(8)}$ of Eq.~(\ref{elf2}).
It coincides with the case without the upper gap boundary
($E_\parallel^{(9)}$ of Eq.~(\ref{app9}), dotted) except for 
$h \mathrel{\hbox{\rlap{\hbox{\lower4pt\hbox{$\sim$}}}\hbox{$>$}}} h_c$.
The dot-dashed line presents $E_\parallel^{(4)}$ of Eq.~(\ref{app4}).
Thin solid line presents the approximation $E_\parallel^{(3)}$ of Eq.~(\ref{app3}).
}
\end{figure}

For $P = 0.1$ s the approximation $E_\parallel^{(9)}$ 
matches the approximation $E_\parallel^{(3)}$ (which is valid
for low altitudes (Eq.~\ref{app3}))
at $h \simeq 0.6r_{\rm pc}$.
This altitude is much lower than estimated by Muslimov \& Harding
(\cite{mh}). 
Consequently, below $\sim 0.6 r_{\rm pc}$ either $E_\parallel^{(3)}$ 
or $E_\parallel^{(4)}$ should be used.

\section{Conclusions}
We find that for  accelerator height $h_c$ approaching the polar cap radius
$r_{\rm pc}$ the accelerating electric field (for nonorthogonal rotators) 
may be approximated according to Eq.~(\ref{app6}):

\begin{equation}
E_\parallel = -1.46\ \frac{B_{12}}{P^{3/2}}\ h \left(1 - \frac{h}{h_c}\right) 
f_1(\xi) \cos\chi\ {\rm Gauss},
\end{equation}
where
$B_{12}=B_{\rm pc}/(10^{12} G)$, $P$ is the rotation period in seconds, $h$
is in cm, for a star with $M=1.4 M_\odot$, $R_{\rm ns}=10^6$ cm.
This result differs from the approximation derived in HM98 
by a considerable period-dependent factor $r_{\rm pc}/R_{\rm ns}$.
Moreover, unlike the HM98' formula it depends on
the magnetic colatitude $\xi$ through the factor $f_1(\xi)$. 
It may be of particular importance for 
a possibility of limiting acceleration by the resonant inverse compton
scattering (see Dyks \& Rudak \cite{dr}).

We emphasize that the corrections presented in this research note 
do not affect
the results of the numerical calculations presented by Harding \& Muslimov
(\cite{hm}) because their numerical procedures include the exact
expressions given by
Eqs.~(\ref{elf1}) and (\ref{elf2}).
However, their analytic estimates 
of the height and width of the acceleration zone and of the maximum particle
energy need to be revised.

Since for an elongated polar gap ($h_c \gg r_{\rm pc}$) 
the upper gap boundary (pair formation front)
influences the electric field only within the negligible ($\ll h_c$) 
upper part of accelerator,
the formulae derived by Muslimov and Tsygan (\cite{mt}) 
for no upper boundary at $h_c$
may be used within the entire gap with no significant overestimates of
electron energies.

\begin{acknowledgements}
This work was supported by KBN grant 2P03D 02117.
We appreciate informative contacts with Alice Harding and Bing Zhang. 
\end{acknowledgements}

\end{document}